\documentclass{PoS}

\title{Lattice Landau gauge quark propagator at finite temperature}

\ShortTitle{Lattice Landau gauge quark propagator at finite temperature}

\author{\speaker{Paulo J. Silva} and Orlando Oliveira \\ %\thanks{A footnote may follow.}\\
       CFisUC, Departamento de F\'{i}sica, Universidade de Coimbra, 3004-516 Coimbra, Portugal \\
        E-mail: \email{psilva@uc.pt}, \email{orlando@fis.uc.pt}}

\abstract{We study the Landau gauge quark propagator, at finite temperature, using quenched lattice simulations. Special focus is given to the behaviour of the momentum space form factors across the confinement-deconfinement phase transition.}

\FullConference{XIII Quark Confinement and the Hadron Spectrum - Confinement2018\\
		31 July - 6 August 2018\\
		Maynooth University, Ireland}

\begin{document}

%====================================================
%====================================================
\section{Introduction and motivation}

The dynamics of QCD generates dynamically a mass for its fundamental quanta. For the quarks the running mass exceeds largely the contribution due to the Higgs mechanism for
small momenta. Indeed, recent lattice simulations at zero temperature of the Landau gauge quark propagator~\cite{Oliveira:2016muq,Oliveira:2018lln}, taking into account only QCD,  show that the running quark 
mass takes values of $\sim 320$ -- $340$ MeV at zero momentum for current quark masses below 10 MeV. However, in a medium where the fundamental
fields feel the effects of temperature and density, as occurs in heavy ion collisions or inside dense stars,
the properties observed at zero temperature are changed. At sufficiently high temperature quarks and gluon become deconfined and a chiral
phase transition is also expected. For example, lattice simulations within pure SU(3) gauge theories show that 
the longitudinal component of the Landau gauge gluon propagator at finite temperature is sensitive to the deconfinement phase transition~\cite{Silva:2013maa}. 
Moreover, due to the breaking of the center symmetry, the propagators in various $Z_3$ sectors differ above $T_c$~\cite{Silva:2016onh} and the differences
can be used to identify the transition to the deconfined regime.
The sign problem prevents lattice simulations from covering the full range of realistic chemical potentials $\mu$ and, therefore, the quark
propagator properties are less studied. 
The two point correlation functions for the gauge sector for two-color QCD at finite density were investigated in~\cite{Hajizadeh:2017ewa} and
some low-momentum gluon screening at high densities was observed.

The properties of quarks and, in particular, its propagators change with $T$ and $\mu$. Herein, we report preliminary results for the study of the Landau gauge quark propagator in momentum space, at finite temperature and for temperatures above and below the deconfinement phase transition, using quenched lattice simulations.
After gauge fixing, we take into account those configurations that belong to the $Z_3$ sector where the phase of the associated Polyakov loop vanishes. 
There are  similar studies where the mass function was measured using Wilson fermions~\cite{Hamada:2006ra}, non-perturbative improved
Clover fermions~\cite{Hamada:2010zz} and where the quark spectral functions were investigated~\cite{Karsch:2009tp,Kaczmarek:2012mb}  relying on particular ans\"atze.
Our study is performed using much larger lattice volumes and looks also to the bare quark mass dependence of the quark propagator form factors.

%====================================================
%====================================================
\section{Quark Propagator at Finite Temperature}

The simulations with finite temperature break rotational invariance and, therefore, in momentum space the continuum quark propagator is described by three
form factors, namely,
\begin{eqnarray}
   S(p_t , \vec{p} ) & = &\frac{1}{ i \gamma_t \, p_t ~ \omega (p_t, \vec{p}) + i \vec{\gamma} \cdot \vec{p} ~ Z(p_t , \vec{p} ) + \sigma (p_t , \vec{p} ) }
                                        \nonumber \\
%\end{equation}
%
%\begin{equation}
%   S(p_t , \vec{p} ) 
& = &\frac{ - i \gamma_t \, p_t ~ \omega (p_t, \vec{p}) - i \vec{\gamma} \cdot \vec{p} ~ Z(p_t , \vec{p} ) + \sigma (p_t , \vec{p} ) }
                                        {   p^2_t ~  \omega^2 (p_t, \vec{p})   + \left( \vec{p}\cdot\vec{p} \right) ~ Z^2(p_t , \vec{p} )  + \sigma^2 (p_t , \vec{p} ) }.
                                        \label{Eq:ContQuarkProp}
\end{eqnarray}
In the results reported below, we use non-perturbative improved Clover fermions~\cite{Sheikholeslami:1985ij} and consider rotated sources~\cite{Heatlie:1990kg}. 
Furthermore, in the analysis of the lattice data we assume that the simulations  are close to the continuum and compute the $\omega (p_t, \vec{p})$, $Z (p_t, \vec{p})$ and $\sigma (p_t, \vec{p})$ 
form factors relying on Eq. (\ref{Eq:ContQuarkProp}). On the lattice given that the boundary conditions for fermions are periodic in
the spatial directions and anti-periodic in time, the momenta $\vec{p}$ and $p_t$ assume discrete values 
\begin{equation}
   p_t = \frac{ 2 \, \pi}{L_t} \, \left( n + \frac{1}{2} \right) \qquad\mbox{ and }\qquad
   p_i = \frac{ 2 \, \pi}{L_s} \,  n \ , \qquad\mbox{where }\qquad n = 0, \, 1, \, 2, \, \dots
\end{equation}
Further, the ``continuum'' form factor is given by $Z_c(p_t, \vec{p}) = Z(p_t, \vec{p}) / \omega (p_t, \vec{p})$ and the running
quark mass by $M (p_t, \vec{p}) = \sigma(p_t, \vec{p}) / \omega(p_t, \vec{p})$.

In the current work, these form factors are computed for temperatures around the critical temperature, taken as the critical temperature above which the gluons
become deconfined for pure Yang-Mills Theory, i.e. $T_c \approx 270$ MeV. The lattice setup used in the simulations is described in Tab.~\ref{Tab:LatSet}, where
each ensemble has one hundred gauge configurations.

The quark propagator is computed inverting the fermionic matrix for 2 point sources. The values used for $c_{sw}$ and the critical hopping parameter $\kappa_{c}$ 
are computed from~\cite{Luscher:1996ug}. In Tab.~\ref{Tab:LatSet} the bare quark mass is given by $m_{bare} = ( 1/ \kappa - 1 / \kappa_c ) / 2 a$, where $a$
is given in~\cite{Silva:2013maa}, and our simulations consider $m_{bare} \approx 10$ MeV or 50 MeV.

\begin{table}
\begin{center}
\begin{tabular}{ccccccc}
\hline
T(MeV)  &  $\beta$  &  $L_s^3 \times L_t$  & $\kappa$ & $m_{bare}$ (MeV) & $c_{sw}$  \\
\hline
243   & 6.0000     &  $64^3\times8$   & 0.1350 & 10   & 1.769\\
      &            &                  & 0.1342 & 53   &      \\
\hline
260   & 6.0347     &  $68^3\times8$   & 0.1351 & 11   & 1.734\\
      &            &                  & 0.1344 & 51   &      \\
\hline
275   & 6.0684     &  $72^3\times8$   & 0.1352 & 12   & 1.704\\
      &            &                  & 0.1345 & 54   &      \\
\hline      
\end{tabular}
\end{center}
\caption{Lattice setup.} \label{Tab:LatSet}
\end{table}

\begin{figure}[t] %  figure placement: here, top, bottom, or page
   \centering
   \includegraphics[width=0.8\textwidth]{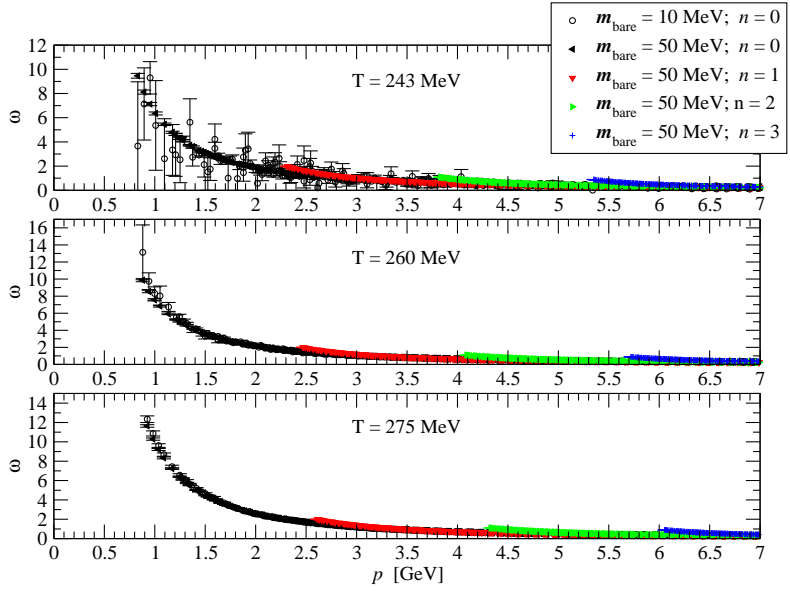} 
   \caption{Bare $\omega$, in lattice units, for the various ensembles and for the various Matsubara frequencies.}
   \label{fig:omega}
\end{figure}

%====================================================
%====================================================
\section{Results and Conclusions}

\begin{figure}[t] %  figure placement: here, top, bottom, or page
   \centering
   \includegraphics[width=3.5in]{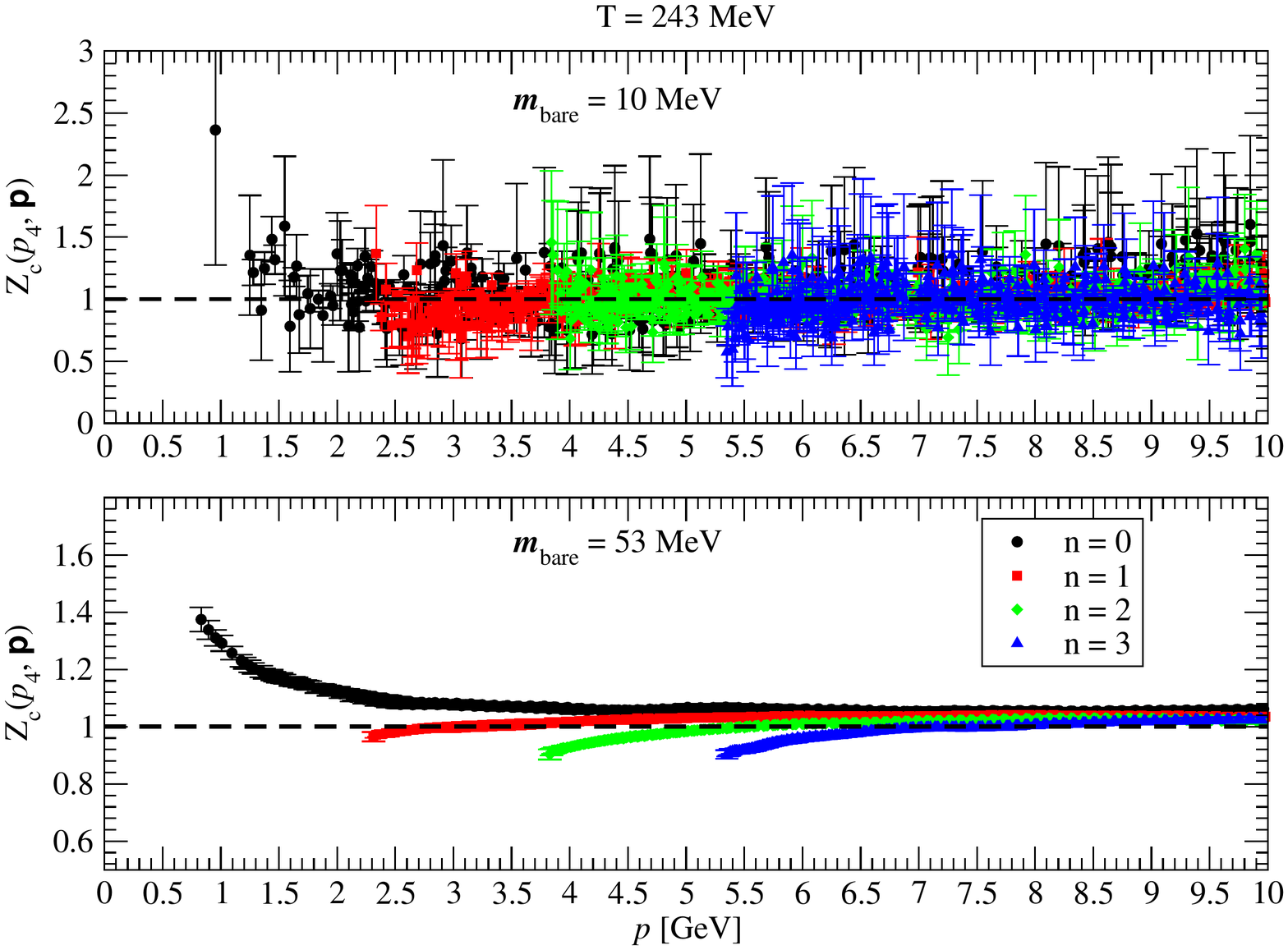} \\
   \vspace{-0.7cm}
   \includegraphics[width=3.5in]{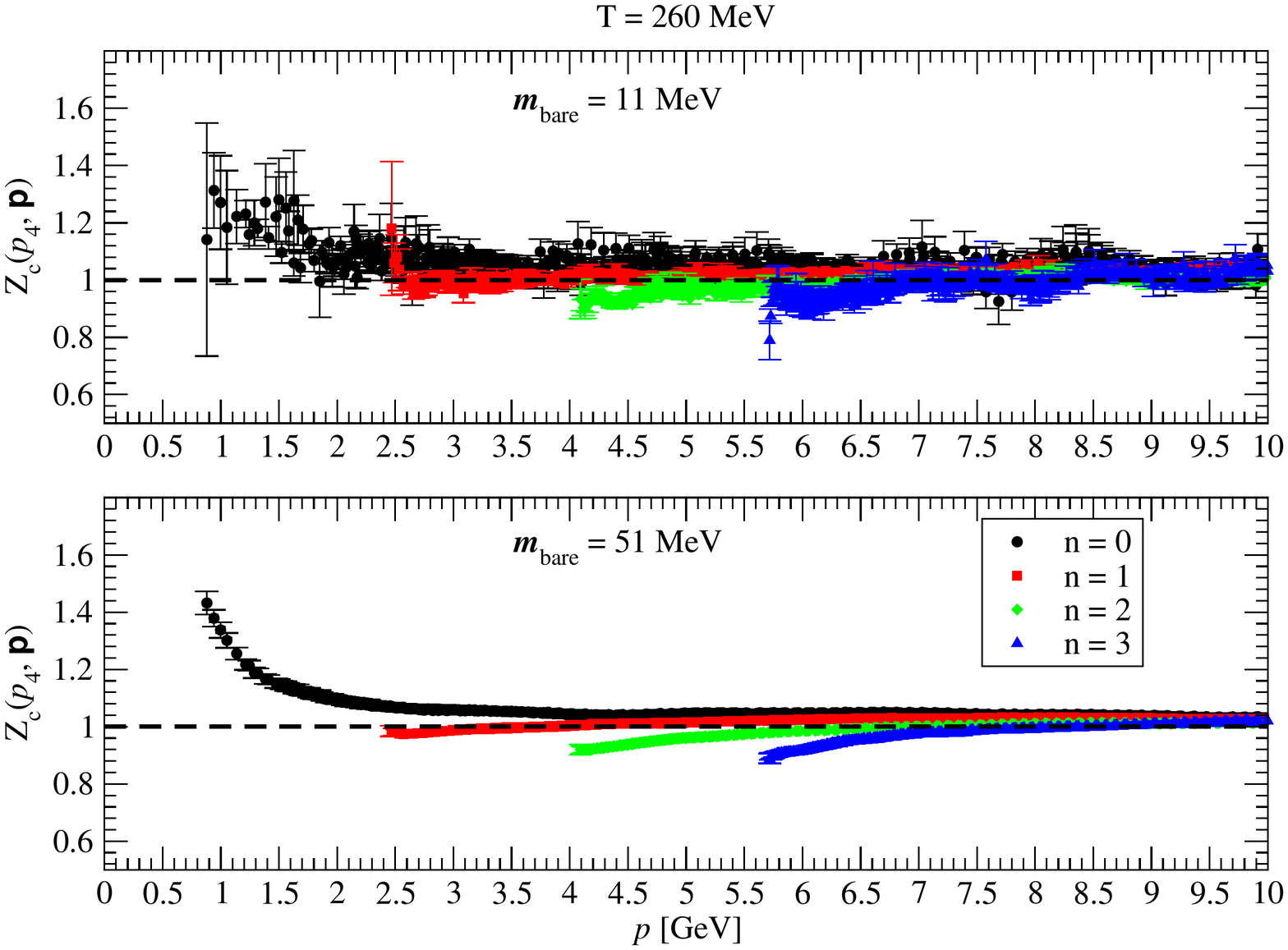} \\
   \vspace{-0.7cm}
   \includegraphics[width=3.5in]{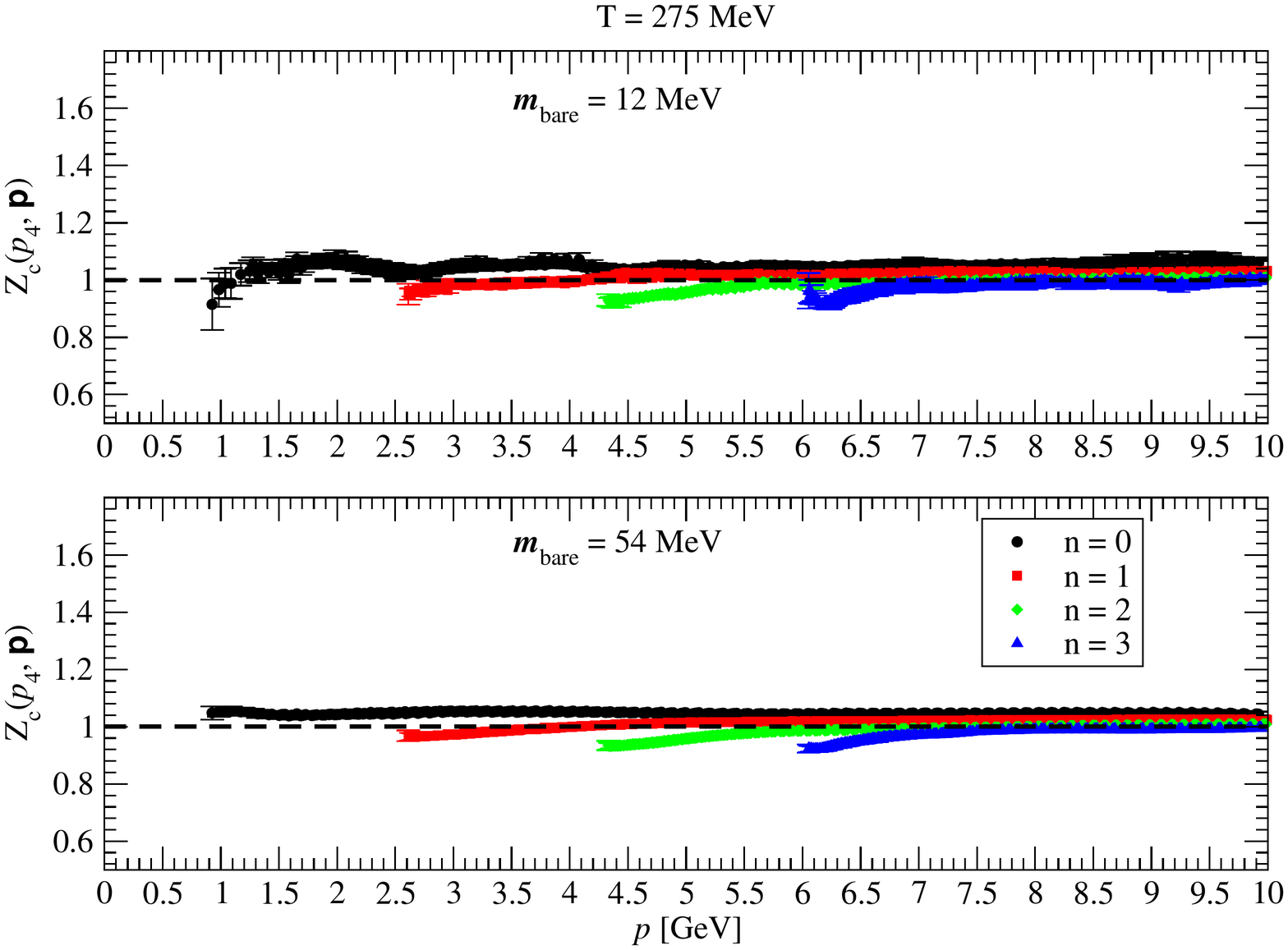}   
   \caption{Bare $Z_c(p_t, \vec{p})$ for the various simulations. }
   \label{fig:Zc}
\end{figure}

In order to suppress lattice effects we have applied momentum cuts as described in~\cite{Aouane:2011fv,Silva:2013maa}. 
For those quantities defined by ratios of lattice functions, as $Z_c$ and $M$, we have ignored all the points whose statistical error was larger than 50\%.

In Fig.~\ref{fig:omega} we report the bare $\omega ( p_t , \vec{p} )$ for all Matsubara frequencies for the heaviest mass considered. For the lightest quark mass only the lowest
Matsubara frequency is shown. The plot shows a general trend observed for all simulations: the function associated with the lightest bare quark mass 
have large fluctuations and, therefore, large statistical errors below the deconfinement phase transition, while above the deconfinement phase transition the statistical fluctuations 
become similar for both quark masses. Furthermore, at sufficiently high momenta no significative difference between the momenta functions associated to the Matsubara
frequencies is seen.

On Fig.~\ref{fig:Zc} the bare $Z_c(p_t , \vec{p})$ is reported. Note that no lattice artefacts are subtracted to the data. 
Below $T_c$, we observe significant statistical fluctuations associated with the lightest quark mass.
 Above the deconfinement temperature the lattice data for the lightest quark mass also shows oscillations that deserve further study.
The function $Z_c(p_t , \vec{p})$ seems to approach a constant for sufficient large momenta. The not so good agreement between the results for the various Matsubara frequencies
at high momenta are a possible indication of the breaking of the O(4) symmetry, the use of finite lattice spacing and/or finite volume effects. 

\begin{figure}[t] %  figure placement: here, top, bottom, or page
   \centering
   \includegraphics[width=3.5in]{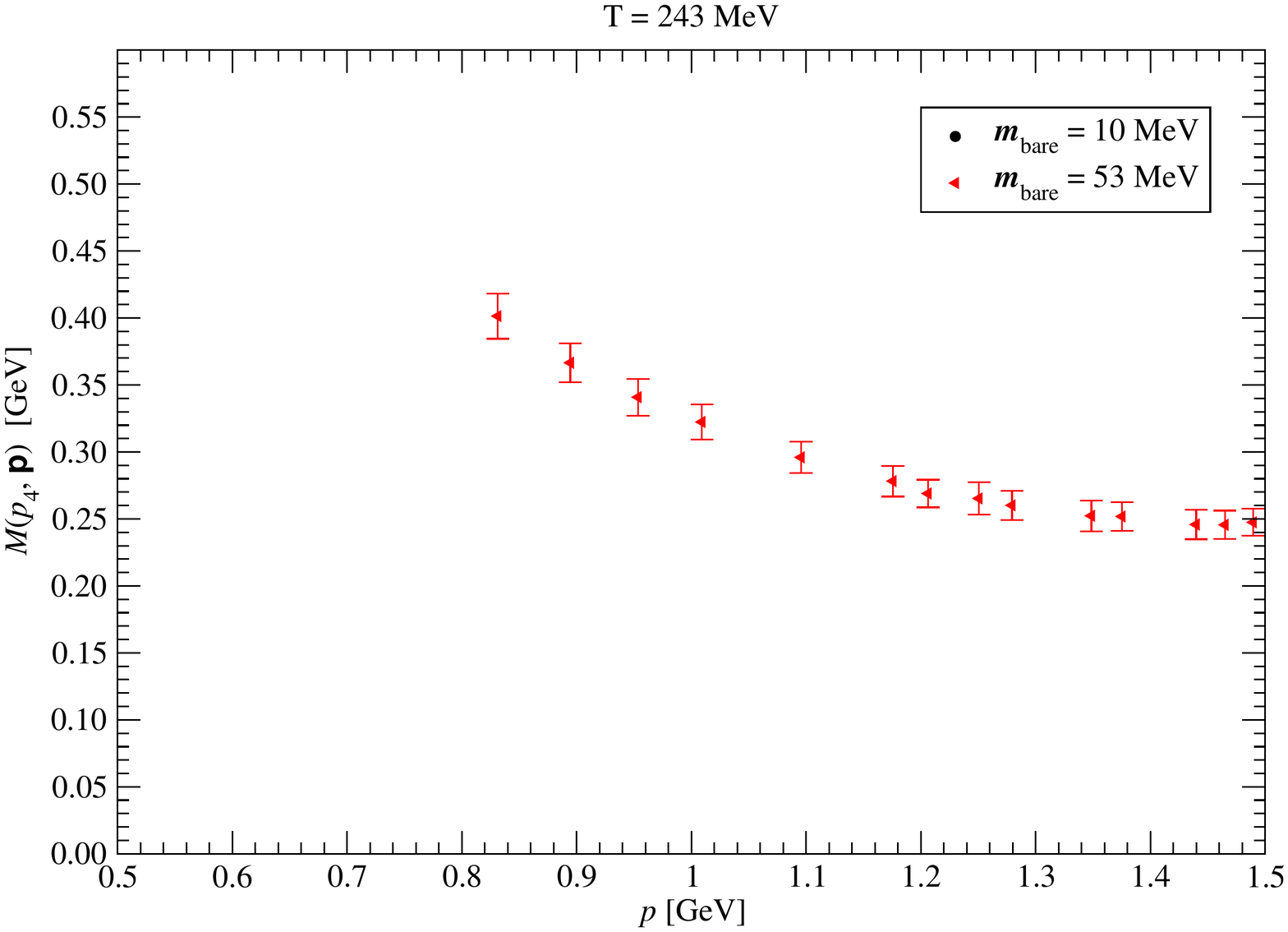} \\
   \vspace{-0.7cm}
   \includegraphics[width=3.5in]{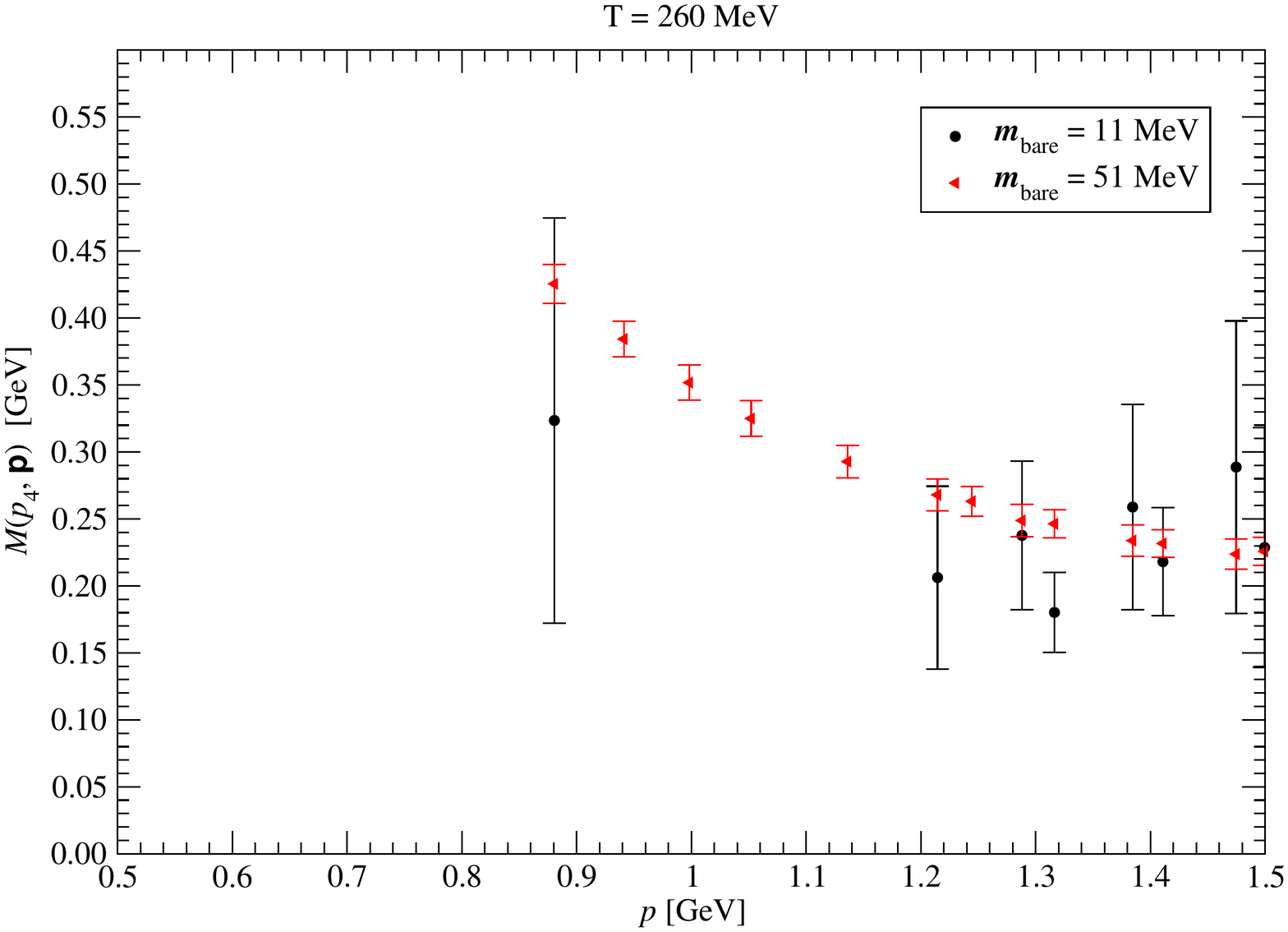} \\
   \vspace{-0.7cm}
   \includegraphics[width=3.5in]{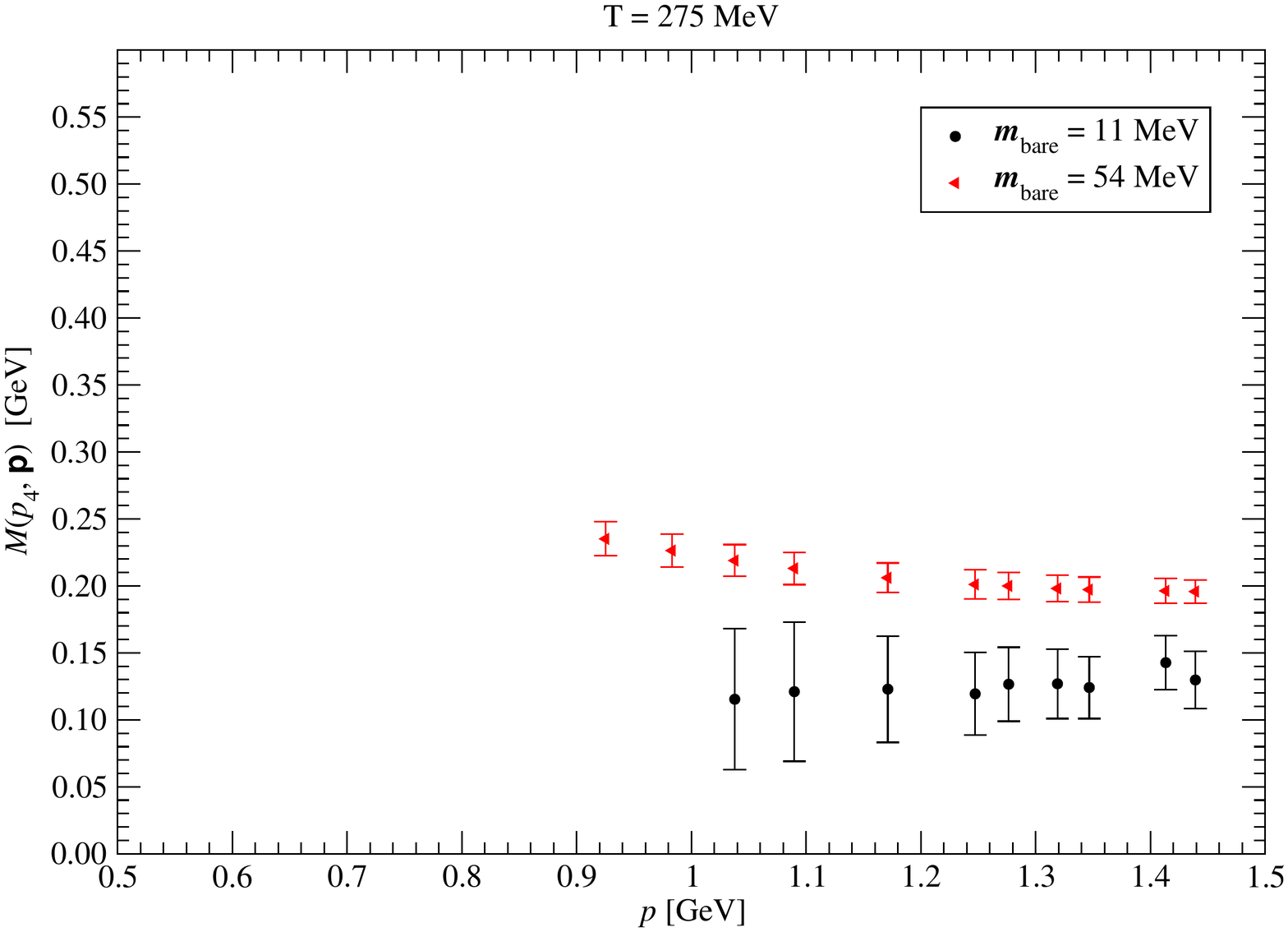}   
   \caption{Running quark mass $M(p_t, \vec{p})$ for the various simulations. }
   \label{fig:M}
\end{figure}

\begin{figure}[t] %  figure placement: here, top, bottom, or page
   \centering
   \includegraphics[width=4in]{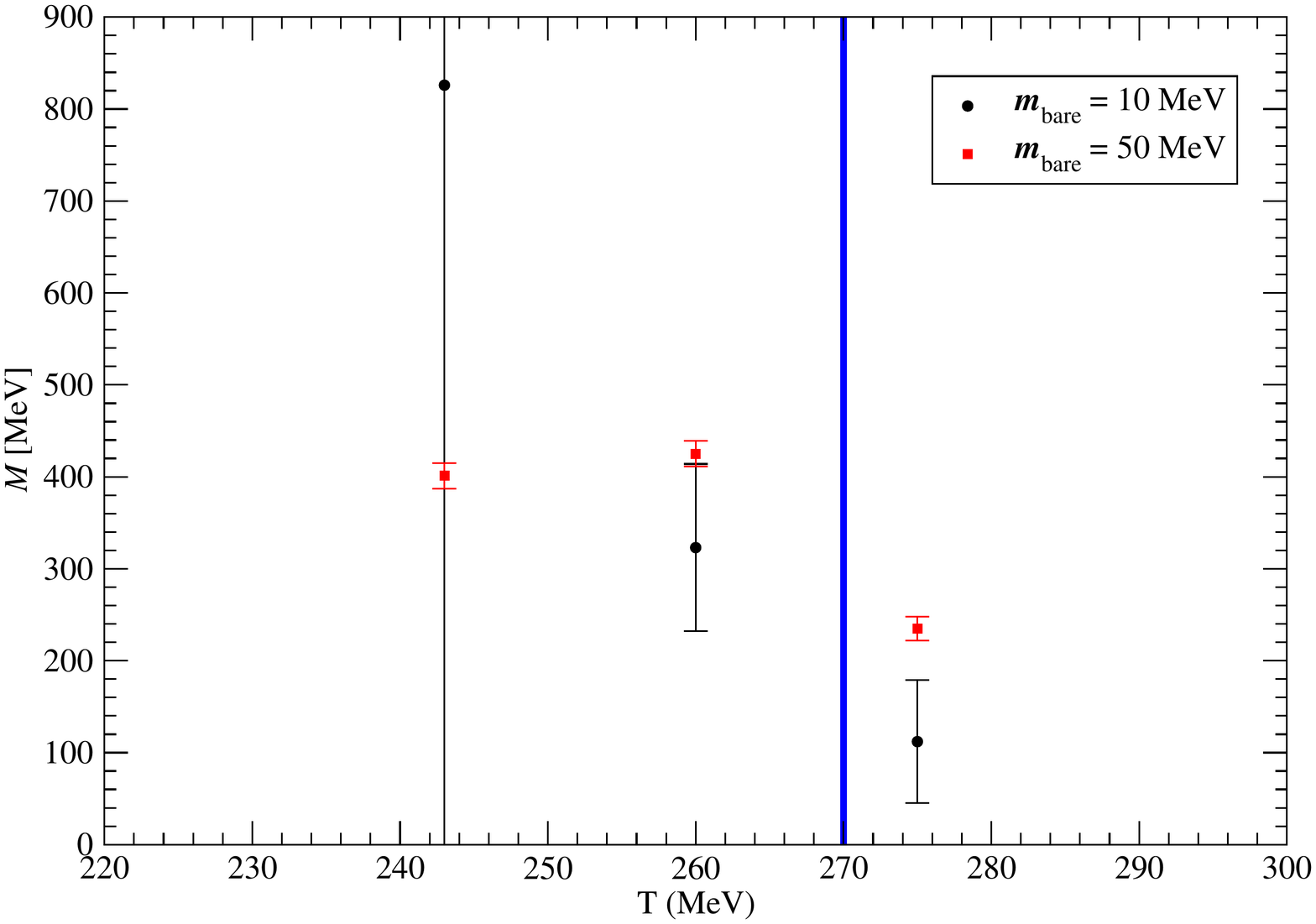} 
   \caption{$M(p_t , \vec{p}=0)$ for the smallest  $p_t$ for the various ensembles. }
   \label{fig:massas}
\end{figure}

On Fig.~\ref{fig:M} we show the running mass for the various simulations. Note that no removal of the lattice artefacts is performed and that why the data is shown
only up to $p = 1.5$ GeV. For higher momenta the corrections to the use of finite lattice spacing are rather large, making the raw lattice data for the running mass
meaningless - see e.g. the discussion and references in~\cite{Oliveira:2018lln}. The data shows a $M(p_4, \vec{p})$ that decreases with the temperature. This effect
is illustrated in Fig.~\ref{fig:massas}, where we report $M(p_t , \vec{p}=0)$ for the smallest  value of $p_t$ for each of the simulations. As we cross the deconfinement
temperature, the blue vertical line on the plot, the value of $M$ becomes about half of its values below $T_c$. Note, however, that $M$ are always much larger than the bare quark masses and, although, the data suggests a decreasing $M(T)$ it can cannot be viewed as an indication of the chiral symmetry restoration.

\section*{Acknowledgements}

P.J.S. acknowlegdes the generous sponsorship by the TUM/LMU "Universe Cluster". P.J.S. also acknowledges support by Funda\c{c}\~ao para a Ci\^encia e a Tecnologia (FCT) under contracts SFRH/BPD/40998/2007 and SFRH/BPD/109971/2015.
The authors acknowledge financial support from FCT under contract with reference UID/FIS/04564/2016.
The authors also acknowledge the Laboratory for Advanced Computing
at University of Coimbra (http://www.uc.pt/lca)
for providing access to the HPC  resource
Navigator. 
The SU(3) lattice simulations were done using Chroma \cite{Edwards2005} and
PFFT \cite{Pippig2013} libraries.

\end{document}